\documentstyle[12pt,epsfig]{article}
\renewcommand{\bar}[1]{\overline{#1}}

\begin{document}
\bigskip\bigskip
\begin{center}
{\large \bf Searching for New Physics \\
by means of Relative Phases \\
and Moduli of Decay Amplitudes}
\end{center}
\vspace{12pt}
\begin{center}
\begin{large}

Z.~J.~Ajaltouni$^{1}$\footnote{ziad@clermont.in2p3.fr},
E.~Di Salvo$^{1,2}$\footnote{Elvio.Disalvo@ge.infn.it}
\end{large} 

\bigskip

$^1$ 
Laboratoire de Physique Corpusculaire de Clermont-Ferrand, \\
IN2P3/CNRS Universit\'e Blaise Pascal, 
F-63177 Aubi\`ere Cedex, France\\

\noindent  
$^2$ 
I.N.F.N. - Sez. Genova,\\
Via Dodecaneso, 33, 16146 Genova, Italy \\  

\noindent  

\vskip 1 true cm
\end{center}
\vspace{1.5cm}

\vspace{10pt}
\begin{center}{\large \bf Abstract}

We illustrate some tests which may shed light on physics beyond the Standard Model. Some of them are especially sensitive to possible signals of new physics. 

\end{center}

\vspace{10pt}

\centerline{PACS numbers: PACS Nos.: 11.30.Er, 11.80.Et, 13.25.-k, 13.30.Eg}

\newpage

\section{Introduction}
Hunt to physics beyond the Standard Model (SM), commonly named new physics (NP), has become particularly lively and intense since the announcement of the LHC start[1-9]. Indeed, although the SM is consistent with a wealth of data\cite{bg}, it is unsatisfactory for several reasons (see [1-6] for recent reviews).

One of the sectors where people look constantly for signals of NP is constituted by processes involving CP violations. Indeed, the Cabibbo-Kobayashi-Maskawa (CKM) scheme, embedded in the SM, is in good agreement with a great deal of present data, however we look for a dynamical description of the effect. In searching for NP, especial attention is being paid to those decays which have shown possible discrepancies with the SM predictions[11-30], stimulating explanations outside the SM and proposals for further investigations in analogous decays. Among them we highlight decays, like $B \to \pi K$[31-38] and $B \to \Phi K^{(*)}$\cite{ba2,be2,ba3}, involving penguin $b \to s$ transitions\cite{hk,dt,dmv,dt2,gr}, to which SUSY and double Higgs models contribute significantly\cite{hk,dt}. 

The possibility of alternative models is also studied in other processes, namely in rare semi-leptonic decays\cite{fke,bsy,bsz,hth,clr}, in neutral meson 
mixings\cite{bug,ksf,bla,bob,pt} and in various hadronic decays[57-73]. These lasts may concern, {\it e. g.}, $U$-spin symmetry violations\cite{dt,nsl}, or the unexpectedly large transverse polarization of vector mesons $V$ in $B\to V ~ V$ decays\cite{dt2,ba11,be11,kp,kp2}. 

Moreover, also time reversal violation (TRV) is used as a probe for NP\cite{alv,cgn,aj}. TRV is commonly regarded as the counterpart of CP violation, in view of the CPT theorem. It is valid under very mild assumptions and not contradicted by any experiments, even supported by stringent tests\cite{pdg}. Up to now direct TRV has been observed only in the CPLEAR experiment\cite{cpl}. Alternatively, TRV may be revealed by the presence, in a hadronic two-body weak decay amplitude, of a "weak" phase, besides the one produced by strong final state interactions (FSI)\cite{va,mrr,bs,wf}.  However, in this case, experimental uncertainties of the "strong" phases create serious problems in singling out the "weak" one\cite{mrr,w2}. Incidentally, this phase occurs in the SM via the CKM scheme.

A more effective tool for investigating possible TRV is provided by triple product correlations\cite{va,wf,bl1,bdl,bdl1,dli}, which may be studied in sequential decays. In fact, comparing a triple product with the CP-conjugated one is sensitive to such violations\cite{dli}. It could even be more effective than CP violations in probing NP\cite{dli}. 

Furthermore, if a sequential decay involves more spinning particles, like $B \to \phi \phi$ or $\Lambda_b \to J/\psi \Lambda$, CP violations, which are usually investigated by means of differences between partial decay widths, could equally well be studied through angular analyses\cite{va}. These studies could be even more convenient\cite{dqs}. Indeed, angular analyses allow the determination of moduli and relative phases of helicity amplitudes\cite{va,ddl,cw,kmp,chg,da}. Thus, in principle, even CPT violations could be detected\cite{chu}.
  
However, until now such observables have not been exploited as an alternative or complementary tool for probing NP. The aim of the present note is to suggest alternative tests for the presence of NP, based on helicity amplitudes, in decays like those mentioned before.

To this end, we derive theoretical consequences of TRV. In this way we propose tests suitable for two-body sequential decays, which can be explored by experiments like those recently suggested or realized for CP violations[11-19,39-41]. 

Section 2 is devoted to deducing the condition for time reversal invariance (TRI) of decay amplitudes. In section 3 we introduce some asymmetries, which allow to apply tests to data. In section 4 we make a few remarks. Finally, our conclusions are presented in section 5.
 
\section{A condition for TRV}

We focus on hadronic two-body decays of the type
\begin{equation}
R_0 ~ \rightarrow ~ R_1 ~~ R_2 \label{dec}.
\end{equation} 
Here $R_0$ is the original resonance and $R_1$ and $R_2$ the decay products, with spin $J$, $s_1$ and $s_2$ respectively.

Our condition for TRV is derived by extending the standard treatment of TRI for two-body decays\cite{mrr,bs} to the case where more than one non-leptonic decay mode is involved\cite{su}. If (\ref{dec}) is a weak decay, the relative, rotationally invariant amplitude reads, at first order in the weak coupling constant,
\begin{equation}
A^J_{\lambda_1 \lambda_2} = \langle f^{out}| H_w |J M\rangle.
\label{wa}
\end{equation}
Here $H_w$ is the weak hamiltonian, $|f^{out}\rangle$ a 
shorthand notation for the final two-body angular momentum eigenstate 
$|J M \lambda_1 \lambda_2\rangle$, $M$ the component of the spin 
of $R_0$ along the $z$-axis of a given frame and $\lambda_1$ and
$\lambda_2$ the helicities of, respectively, $R_1$ and $R_2$
in the rest frame of $R_0$. Assume $H_w$ to be TRI, {\it i. e.}, 
\begin{equation}
TH_wT^\dagger = H_w.
\end{equation}
Here $T$ is the Time Reversal (TR) operator. Then, taking 
into account the antilinear character of $T$ and the rotational invariance
of the amplitude, we get\cite{ms}
\begin{equation}
A^J_{\lambda_1 \lambda_2} = \langle f^{in}| H_w |J M\rangle^*.
\end{equation}
Inserting a complete set of "out" states yields
\begin{equation}
A^{J}_{\lambda_1 \lambda_2} = \sum_n\langle f^{in}|n^{out}\rangle^*
\langle n^{out}| H_w |J M\rangle^*. \label{tr}
\end{equation}
The only terms which survive in this sum correspond to the decay 
modes of $R_0$; furthermore the non-leptonic decay modes give
the main contribution, since they involve a much greater 
coupling constant than the semi-leptonic decay modes. Now we relax the
limitation of the state $|f^{in}\rangle$ to a two-body one;
moreover we express the "out" states in terms of the $S$-matrix.
This is unitary and, owing to TRI\footnote{We neglect 
weak contributions to scattering.}, also symmetric 
with respect to angular momentum eigenstates\cite{ms}. Then eq. 
(\ref{tr}) can be rewritten as
\begin{equation}
A_m = \sum_n S_{mn}A^*_n. \label{trev}
\end{equation} 
Here, omitting spin and helicity indices, 
\begin{equation}
A_m = \langle m^{out}|H_w|R_0\rangle
\end{equation}
and
\begin{equation}
S_{mn} = \langle m^{in}|S|n^{in}\rangle.
\end{equation}
It is worth noting that eq. (\ref{trev}) coincides with eq. (12)
of ref. \cite{su}. The $S$-matrix is block-diagonal\cite{wf}, 
since not all hadronic states are connected to one another by strong interactions. In particular, such blocks are characterized by 
flavor\cite{wf}. 

A solution to eq. (\ref{trev}) can be obtained by diagonalizing 
the $S$-matrix. To this end we recall a result by Suzuki\cite{su}, 
that is, 
\begin{equation}
S_{mn} = \sum_k O_{mk}e^{2i\delta_k}O^T_{kn},  
\end{equation}
where $O$ is an orthogonal matrix and the $\delta_k$ are strong 
phase-shifts. Then the solution to eq. (\ref{trev}) is given by
\begin{equation}
A_m = \sum_n a_n O_{mn} e^{i\delta_n}. \label{sol}
\end{equation}
Here $a_n$ are real amplitudes representing the effects of weak 
interactions on the decay process. Obviously complex values of 
one or more such amplitudes - that is, ``weak'' phases - would 
imply TRV. This result generalizes the one where only elastic scattering 
is allowed between the decay products\cite{mrr}. 

Before considering applications of our model independent result, two important remarks are in order. Firstly, the quantities $O_{mn}$ and $\delta_n$ depend on strong interactions and therefore are invariant under parity 
inversion, P, and under charge conjugation, C. Secondly, if 
a local field theory is assumed - like the SM -, a nontrivial 
phase of at least one $a_n$ implies also CP violation, 
owing to CPT symmetry. In particular, this phase is related to 
the phase of the CKM matrix in the SM. 

\section{Proposed tests for NP}

The problem we are faced with is to determine quantities which are somewhat sensitive to the ``weak'' phases, so as to compare them with SM predictions. Unfortunately, according to the result found in the previous section, we cannot determine those phases. First of all, the elements of the matrix $O$ are not known, at best one can elaborate models, like ref. \cite{su}. Secondly, the amplitudes $ A_n$ may be determined up to a phase per decay mode. In fact, a decay of the type (\ref{dec}), with spinning and unstable decay products, allows to determine, through angular distribution, polarizations and polarization correlations\cite{va,ddl,cw,kmp,chg,da}, all products of the type $A_{\lambda_1\lambda_2}A^*_{\lambda'_1\lambda'_2}$. We stress that such observables may be determined by means of a sequential decay, which is sensitive to non-diagonal elements of the density matrix of the decay products\cite{byf,bsfr,cng}. Moreover we recall a previous paper\cite{da}, where we showed a method for extracting such products from the observables of the decay
\begin{equation}
\Lambda_b \rightarrow \Lambda V. \label{lb}
\end{equation}
These products allow, in turn, to infer all moduli of the amplitudes and their phases relative to a given amplitude, taken as a reference. In this connection, it is worth recalling that recent measurements of sequential decays of beauty resonances to two vector mesons have led to determining moduli and relative phases of decay amplitudes in the transversity representation\cite{be1,cdf,ba2,be2,ba3}, linearly related to the helicity representation. Although such quantities do not allow to determine ``weak'' phases, they may hide signatures for NP, as we shall see in the following subsections. Indeed, we shall present three different types of tests for singling out possible signatures of physics beyond the SM. 

\subsection{Tests for general two-body decays}

Define the following two asymmetries:
\begin{eqnarray}
{\cal A}_{CP} &=& 
\frac{\Phi_{\lambda_1 \lambda_2}-{\bar\Phi}_{-\lambda_1 -\lambda_2}}
{\Phi_{\lambda_1 \lambda_2} +{\bar\Phi}_{-\lambda_1 -\lambda_2}}, \label{asy}
\\
{\cal A}_M &=& \frac{|A^J_{\lambda_1 \lambda_2}|^2-|\bar{A}^J_{-\lambda_1 -\lambda_2}|^2}{|A^J_{\lambda_1 \lambda_2}|^2+|\bar{A}^J_{-\lambda_1 -\lambda_2}|^2}. \label{asy2}
\end{eqnarray}  
Here the barred symbols refer to charge-conjugate amplitudes. Moreover
\begin{eqnarray}
\Phi_{\lambda_1 \lambda_2} &=& tan^{-1}\frac{\Im(A^J_{\lambda_1 
\lambda_2}A^{J*}_{\lambda_{01} \lambda_{02}})}{\Re(A^J_{\lambda_1 
\lambda_2}A^{J*}_{\lambda_{01} \lambda_{02}})} =
sin^{-1}\frac{\Im(A^J_{\lambda_1 \lambda_2}A^{J*}_{\lambda_{01} \lambda_{02}})}{|A^J_{\lambda_1 \lambda_2}| |A^J_{\lambda_{01} \lambda_{02}}|}
\nonumber 
\\
&=& cos^{-1}\frac{\Re(A^J_{\lambda_1 \lambda_2}A^{J*}_{\lambda_{01} 
\lambda_{02}})}{|A^J_{\lambda_1 \lambda_2}| |A^J_{\lambda_{01} \lambda_{02}}|}. 
\label{phas}
\end{eqnarray}
are relative phases; $\lambda_1$ and $\lambda_2$ are the helicities of the decay products in the overall center-of-mass, as introduced in sect. 2. Lastly 
$A^J_{\lambda_{01}\lambda_{02}}$ denotes the reference amplitude.
If at least one of the amplitudes $a_n$ is complex, the above asymmetries, (\ref{asy}) and (\ref{asy2}), will be generally nonzero.
In particular, this is true if CPT symmetry is assumed. 

The experimental values of such asymmetries - analogous to observables proposed by other authors\cite{dli} - have to be compared to SM predictions. In any case, if ${\cal A}_{CP}$ or ${\cal A}_M$ is nonzero, we may conclude that CP is violated, but not necessarily TR. 

\subsection{Tests for particular decays}

A particularly intriguing case is represented by two-body decays that fulfil the condition 
\begin{equation}
{\cal A}_M = 0. \label{cond}
\end{equation}
This corresponds to one of the following two situations:

~i) the sum (\ref{sol}) consists of just one term,

ii) or the ``weak'' amplitude $a_n$ (or at least its phase\footnote{In the following we do not consider this case, which appears quite unrelistic.}) can be factored out of that sum.

The latter case is typical of a unique basic quark process contributing to all decay modes of a given block of the $S$-matrix\cite{wf,su}. For example, some of the $b\to s$ transitions are dominated in the SM by the penguin 
diagram\cite{hk,dt,dmv,dt2,gr}, the tree diagram being forbidden. In this case the sum (\ref{sol}) amounts to  
\begin{equation}
A_m = a \sum_n O_{mn} e^{i\delta_n}, \label{sol1}
\end{equation}
$a$ being the common ``weak'' amplitude. The sum in eq. (\ref{sol1}) concerns only strong interactions and is related to the absorptive part of the amplitude of a weak decay to intermediate states, followed by strong interactions that lead to the final state\cite{wf}. To be precise, condition (\ref{cond}) does not necessarily imply eq. (\ref{sol1}), however it is rather unlikely that, with different ``weak'' amplitudes $a_n$ in the sum (\ref{sol}), the amplitude $A_m$ and its CP-conjugated have the same modulus square. Of course this is not true if CPT symmetry is assumed. Then, if condition (\ref{cond}) is satisfied, we assume that the decay is driven by a single dominant diagram, for example penguin or tree. Among penguin dominated decays, especially interesting for NP are the following decay modes of $B$, $B_s$ and $\Lambda_b$, in part already studied, either theoretically\cite{dt,dmv,dt2,gr} or experimentally\cite{ba2,be2,ba3}:

\begin{eqnarray}
B &\rightarrow& \phi K^{*}, ~~~ K^{*} {\bar K}^{*}; \ ~~~ ~~~~~~~~~~~~~ \
\label{b+}
\\
B_s &\rightarrow& \phi \phi, ~~~ J/\psi {\bar K^{*0}},
\label{b0}
\\
\Lambda_b &\rightarrow& \Lambda \phi. ~~~~~~~~~~~~~ \ ~~~ ~~~~~~~~~~~~~ \
\label{lbv}
\end{eqnarray}

These decays are dominated by the penguin diagram, driven by the subprocess 
\begin{equation}
b\to s + {\bar q} q. \label{qp}
\end{equation}
Under condition (\ref{cond}), it makes sense to define also the following two asymmetries: 
\begin{eqnarray}
{\cal A}_{C} &=& \frac{\Phi_{\lambda_1 \lambda_2} -{\bar\Phi}_{\lambda_1\lambda_2}}
{\Phi_{\lambda_1 \lambda_2} +{\bar\Phi}_{\lambda_1\lambda_2}},\label{asy3} 
\\
{\cal A}_{P} &=& \frac{\Phi_{\lambda_1 \lambda_2} -{\Phi}_{-\lambda_1 -\lambda_2}} {\Phi_{\lambda_1 \lambda_2} +{\Phi}_{-\lambda_1 -\lambda_2}}.
~~~~~~~~~~~~ ~~~~~~~~~~~~ ~~~~~~~~~~~~ ~~~~~~~~~~~~ \label{asy4}
\end{eqnarray}  
If condition (\ref{cond}) is fulfilled, and simultaneously at least one of the asymmetries (\ref{asy}), (\ref{asy3}) and (\ref{asy4}) is different from zero, we may conclude that traces of NP are evident. Indeed, according to the CKM scheme, the ``weak'' phase is independent of hadron helicities, therefore the asymmetries defined above are predicted to be zero. A nonzero value of one of such asymmetries would be a signature of NP. Helicity dependent ``weak'' phases could be realized, for example, in the case of flavor changing neutral currents, by interference between $Z$ (or $Z'$) exchange and Higgs exchange: the former selects only left-handed quarks, while the latter accepts equally left- and right-handed ones. 

Our method can help to resolve ambiguities illustrated in ref. \cite{imlt}, about singling out NP contributions in penguin diagrams.

Aside from that, an indication of NP may come from a decay mode which, according
to the SM, is governed by a single (penguin or tree) graph and yet it does not satisfy condition (\ref{cond}). A somewhat analogous procedure was suggested by
Datta and London\cite{dli}. In this connection we recall that interference between a tree diagram and a NP contribution, if present, would be probably detected\cite{ikm} thanks to the new facilities. 

\subsection{Interference between two ``weak'' amplitudes}

Different, but equally interesting cases are constituted by the decay modes which according to the SM are driven by interference between the tree and the penguin diagram\cite{ba1,ffm,dt,ddl,kp,kp2}. Typical decays of this type are given by 
\begin{eqnarray}
B &\rightarrow& J/\psi K^{*}, ~~~ \rho K^{*}, ~~~ \omega K^{*}; ~~~~~~~~~~ \
\label{bb+}
\\
B_s &\rightarrow& J/\psi \phi, ~~~ J/\psi {\bar K}^{*0}; ~~~~~~~~~~ \
~~~~~~~~~~ \
 \label{bb0}
\\
\Lambda_b &\rightarrow& \Lambda J/\psi, ~~~~~ \Lambda \rho, ~~~~~ \Lambda \omega. ~~~~~~~~~~ \ \label{lbbv}
\end{eqnarray}
In these cases eq. (\ref{sol1}) has to be replaced by
\begin{equation}
A_m = a_t T_m + a_p P_m, \label{sol2}
\end{equation}
where $a_{t(p)}$ is the ``weak'' amplitude for the tree (penguin) diagram and   
\begin{equation}
T_m(P_m) = \sum_{n_{t(p)}} O_{mn_{t(p)}} e^{i\delta_{n_{t(p)}}}
\end{equation} 
describes the effects of the strong interactions in each such diagram. Here 
$n_{t(p)}$ runs over the intermediate strong states available in a decay described by a tree (penguin) diagram. The two spectra do not coincide, but may have some overlap. Among the decays above, those which involve the $J/\psi$ resonance in the final state are dominated by the two quark sub-processes (\ref{qp}) and 
\begin{equation}
b \to c + \bar{c} s,
\end{equation} 
the latter corresponding to the tree diagram.
The asymmetry (\ref{asy2}) is expected to be different from zero, its size giving information on the relative weight of the two amplitudes. If, as in the case of decay (\ref{bb0}), the penguin contribution is very weak, one can apply the tests described in the previous subsection: if ${\cal A}_M$ is quite small and at least one of the other asymmetries, concerning relative phases, is significantly different from zero, this is an indication of NP. 

A further test, similar to the one suggested in subsection 3.2, can be stated by following a procedure similar to the one suggested in ref. \cite{cw}. In that article the authors propose to infer, from angular distribution and polarization of $B\to V V$ decays, the relative ``weak'' phase of the penguin amplitude to the tree one, assuming that such a phase is independent of vector meson helicities. If this assumption is not confirmed by data, we may conclude that NP traces are present.    

\section{Remarks}

At this point some remarks and comments are in order. 

A) We have already observed at the beginning of sect. 3 that recent data yield
moduli and relative phases of decay amplitudes. Our tests may be applied,
provided data become available for the antiparticle decays as well.
 
B) Such data have shown finite relative phases\cite{be1,cdf,ba2,be2,ba3}, which implies sizeable ``strong '' phases produced by FSI, in qualitative agreement with the estimations by Wolfenstein\cite{w2}. These phases may be greater than the ``weak'' ones, which may be an obstacle in the detection of TRV. However, this difficulty is overcome by our method: as recalled at the end of sect. 2, the ``strong '' phases are eliminated in the numerators of asymmetries (\ref{asy}), owing to C and P invariance of strong interactions. Similar suggestions were given by other authors\cite{fay,va,wf,bl1,bdl,dli}.

C) The application of our method demands a particular care in decays of neutral, spinless resonances to neutral particles, like some of the $B^0$ and $B_s$ decays mentioned in the previous subsection, especially if the decay products constitute a CP eigenstate\cite{ddl,dqs}. In such cases particle-antiparticle mixing complicates the extraction of the parameters needed for our tests. A useful tool for splitting the angular distribution into CP-even and CP-odd contributions is the transversity representation\cite{dqs}. Moreover a suitable quantity for revealing CP violations, and possibly NP, is defined in ref. \cite{dli}.

D) The condition of vanishing of the asymmetry $A_M$, which if fulfilled could reveal clear signatures of NP, is opposite to the one demanded for detecting CP violations, which needs interference among amplitudes, and therefore at least two different weak decay amplitudes of the type $a_n$ in the sum (\ref{sol}).

\section{Conclusions}

We have illustrated a theoretical condition for TRV in non-leptonic two-body decays. This suggests some kinds of tests for possible probes of NP, to be applied to decays to unstable, spinning decay products. The tests may be realized by means of standard analyses. We propose to apply them particularly to those decay modes for which hints of NP have been already found[11-30], or/and suggestions for new investigations have been given[20-38]. Especially effective appear the tests proposed in subsections 3.2 and 3.3, since nonzero values of asymmetries would imply NP. 

Obviously, the feasibility of the tests proposed is conditioned by statistics. Such tests appear to be not so unrealistic, given the wealth of $b-{\bar b}$ pairs to be produced at LHC per year ($10^{12}$). The asymmetries proposed do not derive significant contribution from the SM; but, if NP is present, and it is realized according to some of the alternative models, it could be detected.  

\vskip 0.25in
\centerline{\bf Acknowledgments}
The authors are grateful to their friend J. Orloff for stimulating
discussions. One of us (EDS) is also indebted to his friend A.
Di Giacomo for useful suggestions.

\end{document}